\documentstyle[epsfig,12pt]{article}
\input epsf
\begin{document}
\bibliographystyle{unsrt}
\def\question#1{{{\marginpar{\small \sc #1}}}}
\newcommand{\bra}[1]{\left < \halfthin #1 \right |\halfthin}
\newcommand{\ket}[1]{\left | \halfthin #1 \halfthin \right >}
\newcommand{\be}{\begin{equation}}
\newcommand{\ee}{\end{equation}}
\newcommand{\vsig}{\mbox {\boldmath $\sigma$\unboldmath}}
\newcommand{\vep}{\mbox {\boldmath $\epsilon$\unboldmath}}
\newcommand{\fn}{\frac 1{E^i+M_N}}
\newcommand{\fs}{\frac 1{E^f+M_N}}
\newcommand{\qqbar}{$q \bar{q}~$}
\newcommand{\gsi}{\,\raisebox{-0.13cm}{$\stackrel{\textstyle>}
{\textstyle\sim}$}\,}
\newcommand{\lsi}{\,\raisebox{-0.13cm}{$\stackrel{\textstyle<}
{\textstyle\sim}$}\,} 

\rightline{hep-ph/9710450}
\rightline{RAL-97-048}
\baselineskip=18pt
\begin{center}
{\bf \LARGE 
Filtering glueball from $q\bar{q}$ production in proton proton or 
double tagged  $e^+e^- \rightarrow e^+e^-R$ and implications for 
the spin structure of the Pomeron. }\\
\vspace*{0.9in}
{\large Frank E. Close}\footnote{\tt{e-mail: fec@v2.rl.ac.uk}} \\ 
\vspace{.1in}
{\it Rutherford Appleton Laboratory}\\
{\it Chilton, Didcot, OX11 0QX, England}\\ 
\end{center}

\begin{abstract}

The production of $J^{PC} = 1^{++},0^{-+}$ and $2^{-+}$ mesons in double tagged 
$e^+e^- \rightarrow e^+e^-R$ is calculated and found to have the 
same polarisation and dynamical characteristics as observed in 
$pp \rightarrow ppR$. Implications for the spin structure of the 
Pomeron are considered. Production of $0^{++},2^{++}$ mesons in 
these two processes may enable the dynamical nature of these 
mesons to be determined.

\end{abstract}

\hspace*{2em}

\newpage

Recently it has been discovered that the pattern of resonances 
produced in the central region of double tagged $pp \rightarrow
 p+p+R$ depends on the vector {\it difference} of the transverse
momentum recoil of
 the final state protons\cite{ck97,wa102} 
(even  at fixed four-momentum transfers
$t \sim -k_{1T}^2, t' \sim -k_{2T}^2$, see fig. 1 for kinematic definitions).
 When this quantity 
 ($dk_T \equiv |\vec{k}_{T1} - \vec{k}_{T2}|$)  is small,
($\leq O(\Lambda_{QCD})$), all well 
established
 $q\bar{q}$ states are observed to be suppressed while the 
surviving resonances include
 the enigmatic  $f_0(1500), f_J(1710)$ which have been
proposed as glueball candidates\cite{cafe}.
 At large $dk_T$, by contrast, $q\bar{q}$ states are prominent,
 there appearing to be some correlation between their prominence
 and the internal angular momentum of their $q\bar{q}$ system
 such that high $L$ states turn on more with increasing $dk_T$
 than do their low $L$ counterparts\cite{kirkhad97}. It has been 
suggested that
 this might form the basis of a glueball - $q\bar{q}$ filter since
 $0^{++},2^{++}$ glueballs need no internal angular momentum in
 contrast to the analogous $q\bar{q}$ $^3P_{0,2}$ combinations\cite{ck97}
and the dynamics may thereby favour glueballs as $dk_T \to 0$.

In order to gain insight, we have computed the production in a simple
model where high energy $pp$ interactions are mediated by a preformed
colour singlet object that couples to the proton $\sim \gamma_{\mu}$\cite{lp}.
We find that when the resonance, $R$, has
$J^{PC} = 1^{++}$,
$0^{-+}$ or $2^{-+}$ the predicted $dk_T$ dependence appears to be identical to 
that empirically observed in $pp \rightarrow 
p+p+R$ implying that central production of resonances is mediated
by conserved vector currents independent of the nature of the
meson, $R$. In contrast, we find that for
 $J^{PC} = (0,2)^{++}$ the structure of $R$ can 
be important enabling in principle a filtering of $q\bar{q}$ from glueballs
to be realisable.

To be explicit, we shall calculate the 
production rate and momentum dependences of 
$\sigma(e^+(p_1)e^-(p_2)  \rightarrow e^+(p_3) + e^-(p_4)+ R)$ as
this is well defined in QED and shares topological similarities to
the hadronic processes of interest.
First we shall generalise Cahn's analysis 
of single tagged $e^+e^-$ \cite{cahn} ($\gamma^* \gamma \to R$)
 to the double tagged 
case ($\gamma^* \gamma^* \to R$)
for a general $1^{++}$ state. We define the production 
amplitude

\begin{eqnarray}
{\cal M}=e^2 \bar{u}(p_3) \not{\epsilon^*} u(p_1) \bar{u}(p_4) \not{\epsilon'^*}
u(p_2)
\frac{1}{k_1^2k_2^2} \epsilon_{\mu}T^{\mu \nu}\epsilon'_{\nu}
\end{eqnarray}
where for $1^{++}$ production the $\epsilon_{\mu}T^{\mu \nu}\epsilon'_{\nu}$
may be 
written\cite{cfl,kuhn} (with $k_1 \equiv p_1 - p_3$ and $k_2 \equiv
p_2 - p_4$)

\begin{eqnarray}\label{tmunu}
\epsilon_{\mu}T^{\mu \nu}\epsilon'_{\nu}
[1^{++}]= & A_1(k_1;k_2) \epsilon_{\mu \nu \alpha \beta}
\xi^{\beta} (G_1^{\mu \nu} G_2^{\alpha \delta} k_{\delta2}
+G_2^{\mu \nu} G_1^{\alpha \delta} k_{\delta1}) \nonumber \\
& + A_2(k_1;k_2)\epsilon_{\mu \nu \alpha \beta} \xi_{\delta}
(k_1^{\delta} - k_2^{\delta})G_1^{\mu \nu} G_2^{\alpha \beta}
\end{eqnarray}
where we use
the shorthand $G_{\mu\nu} \leftrightarrow
k_{\mu} \epsilon_{\nu} - \epsilon_{\mu} k_{\nu}$ and the convention
that $G_{1(2)}$ refers 
to $k_{1(2)},\epsilon_{1(2)}$; the $A_i(k_1;k_2)$ are form factors 
to be determined experimentally and $\xi$ is the spin polarisation
vector for the axial meson.

For the special case of non-relativistic $^3P_1Q\bar{Q}$ one 
has\cite{cfl,kuhn} $A_2 \equiv 0$. It is straightforward to verify that 
the tensors multiplying $A_1$ may be
written $ \equiv 
\epsilon_{\mu \nu \alpha \beta} \xi^{\beta} \epsilon(1)^{\nu}
\epsilon(2)^{\alpha} (k_1^2 k_2^{\mu} - k_2^2 k_1^{\mu})$
 as in Cahn's 
eqn(A1). In this case the double tagged differential cross section is

\begin{eqnarray}\label{xsection}
\frac{d\sigma}{dxdydtdt'd\phi} \equiv \frac{e^2}{512 \pi^4 s}
\frac{|A_1(t,t')|^2}{t^2t'^2} \delta(m^2+P_T^2 - xys -(x+y)(t+t')) |M|^2
\end{eqnarray}
where
\begin{eqnarray}\label{rate1}
|M|^2 \equiv 2tt'[t'(su+s'u') + t(su'+s'u) - 2 cos \phi \sqrt{tt'uu'}
(s+s' - \frac{m^2}{2})] \nonumber \\
+
\frac{(t+t')^2}{2m^2}[8uu'tt'sin^2\phi + (s-s')^2
+ (u-u')^2]
\end{eqnarray}
which reduces to Cahn's eqns (A18,19,23) as $t \rightarrow 0$. Here
$s,t,u$ are standard and $s'=2p_3 \cdot p_4; t'=-2p_2 \cdot p_4;
u'=-2p_2 \cdot p_3$
which are related to the mass $m$ of the resonance by
$s+s'+t+t'+u+u' = m^2$ and$P_T$ is the recoil transverse momentum
of the produced resonance.
Furthermore we define 
$cos\phi \equiv \hat{p}_{3T}\cdot \hat{p}_{4T}$
when $p_{1,2}$ are aligned along the $\hat{z}$ axis; (note for 
future reference that the $dk_T$ phenomenon observed in $pp 
\rightarrow p+p+R$ is equivalent to a $\phi$ dependence of 
the cross section). We can make contact with the formalism used 
in ref\cite{ck97} by defining $x,y$ as the fractional energy loss of 
the beams such that
$p_3^z \approx (1-x)p_1^z, p_4^z \approx (1-y)p_2^z$. Then eqn. \ref{rate1}
may be written in the symmetric form

\vskip 0.2in

\begin{eqnarray}\label{xy}
\frac{|M|^2}{-2tt's^2} = & \nonumber \\
& t'(1-y)[1+(1-x)^2]+ t(1-x)[1+(1-y)^2] \nonumber \\
& +2 cos \phi  \sqrt{tt'(1-x)(1-y)}
[1+ (1-x)(1-y) -\frac{xy}{2}] \nonumber \\
& -
\frac{(t+t')^2}{4m^2}([1+(1-y)^2][1+(1-x)^2] - 4(1-x)(1-y)cos2\phi)
\end{eqnarray}

Consider now the particular limit, analogous to that in the $pp$ 
process, $t,t' << m^2$. Writing $t(1-x) = 
-k_{T1}^2$ and $t'(1-y) = -k_{T2}^2$  and then taking the 
limit $x,y \rightarrow 0(\frac{1}{\sqrt{s}})$,  eqn\ref{xy} collapses to 

$$
|M|^2 = 4tt's^2 \times [|\vec{k}_{T1}- \vec{k}_{T2}|^2] \equiv 4tt's^2|dk_T|^2.
$$
Hence as $dk_T \rightarrow 0$ we 
predict that 

\begin{eqnarray}
Lim(dk_T \rightarrow 0) \frac{d\sigma}{dk_T}
(e^+e^- \rightarrow e^+(k_{T1})e^-(k_{T2})R) \rightarrow 0.
\label{dkt}
\end{eqnarray}
We note that this is the same phenomenon observed in the 
$pp$ analogue\cite{ck97,wa102}.

We find also that the $1^{++}$ should be spin polarised in the
$\gamma^* \gamma^*$ c.m. frame. Following the approach of 
ref.\cite{bcl} we predict

\begin{eqnarray}
\frac{\sigma(\lambda=0)}{\sigma(\lambda=\pm 1)} 
=  \frac{2(t-t')^2}{(t+t')m^2 - 2tt'}
\label{pol}
\end{eqnarray}
such that when $t,t'<<m^2$ the $1^{++}$ $Q\bar{Q}$ will be 
produced dominantly with $\lambda=\pm1$. Here again, the 
phenomenon predicted for $e^+e^-$ is apparently manifested 
empirically in 
$pp \rightarrow ppR$\cite{kirkhad97,gaston}, suggesting that
the production is driven by conserved vector currents. 

The suppression of $1^{++}$ as $dk_T \rightarrow 0$ is more 
general than for the specific $Q\bar{Q}$ case considered above. 
Inspection of the general amplitude, eqn (\ref{tmunu}), shows that 
the tensor multiplying $A_2$ vanishes 
as $k_{1T} - k_{2T} \rightarrow 0$. The production rate
therefore also vanishes even when $A_2(k_1;k_2) \neq 0$  (assuming there is no 
pathological singularity in the $A_2$ form factor). 
Hence vanishing of axial meson production in this 
kinematic configuration is general for any production mechanism 
driven by conserved vector currents. 

The similarity in behaviour between that observed in $pp 
\rightarrow p+p+R$ and that predicted in the analogous $e^+e^-$
arises for $R \equiv 0^{-+}$ too. 
 As noted by Castodi and Fr\`ere \cite{cast}
the production of $0^{-+}$ will naturally vanish 
as $dk_T \to 0$ if it is due to conserved vector current
exchanges since in this case the
production amplitude 
is proportional to 

\begin{equation}\label{pseud}
\epsilon_{\mu}T^{\mu \nu}\epsilon'_{\nu}
[0^{-+}] \equiv P(k_1;k_2) \epsilon^{\mu \nu 
\alpha \beta} G_{\mu \nu} G_{\alpha \beta}
\end{equation}
which may be rewritten, in the meson rest frame,as 
$2MP(k_1,k_2) \epsilon_{ijk} (k_1 - k_2)_i \epsilon(1)_{j} 
\epsilon(2)_{k}$. Hence in the absence of a singular form factor 
this 
will vanish as $k_{1T} - k_{2T} \rightarrow 0$. The data of 
ref\cite{wa102,kirkhad97} exhibit such a behaviour in
$pp \rightarrow pp+ \eta(\eta')$. For non-relativistic $Q\bar{Q}$
spin singlets, ($0^{-+},2^{-+}$etc), where
the production amplitude is proportional to derivatives of the wavefunction,
the above structure (eq.\ref{pseud}) 
will be generic (e.g. $2^{-+}$ in ref.\cite{cahn91}).
Hence this sequence should disappear as $k_{1T} - k_{2T} \rightarrow 0$.
This also is found to be true empirically for the $\eta_2(1620)$ and
$\eta_2(1875)$ in $pp \rightarrow pp+\eta_2$\cite{kirkhad97,wa4pi}.

>From the above analysis we infer that the $dk_T \rightarrow 0$ suppression
for $0^{-+}$,
$2^{-+}$ and $1^{++}$ production and the polarisation of the $1^{++}$
will all arise if the initiating fields are conserved vector currents.
Thus they will naturally occur
in $pp \rightarrow ppR$ if the resonance production 
is driven by 
conserved vectors, e.g.
if the pomeron acts as a single hard gluon with colour neutralisation
even at small $t$ (comapre and contrast ref\cite{white}) and that
production of $q\bar{q}$ is via gluon - 
gluon fusion. This is suggestive though not a proof. However, it can
already be concluded from the WA102 phenomenon (refs.\cite{ck97,wa102})
 that the 
pomeron must have a non-trivial helicity structure in order to 
generate non-trivial $\phi$ dependence\cite{landnacht}. Thus the 
Pomeron cannot be simply a scalar or pseudoscalar, in contrast to 
some outdated folklore, nor can it 
transform as simply the longitudinal component of a (non-
conserved) vector\cite{landnacht,landdon}.
The implications of the Donnachie-Landshoff pomeron
for $\phi$ dependence in central production merit study as do
the general implications of $\phi$
dependence for the spin content of the Pomeron. 

For the
particular case of $gg$ fusion, or for $\gamma \gamma$ production, we
may generalise the above analyses to $0^{++}$ and $2^{++}$ following
refs.\cite{cfl,kuhn,ren}.

  A linearly independent set of Lorentz and gauge
invariant production amplitudes 
for $J^{++}$ states is given in \cite{kuhn,ren}.The forms
for $0^{-+}$ and $1^{++}$ in eqns.\ref{tmunu} and \ref{pseud} are as defined in
ref.\cite{kuhn,cfl}. The $0^{++}$ and $2^{++}$ cases are written
\begin{eqnarray}
\epsilon_{\mu}T^{\mu \nu}\epsilon'_{\nu}
[0^{++}] & = &  \frac
{P_{\rho\sigma}}{\sqrt{3}}  \left [
S_1(k_1,k_2)G_{\mu\rho}^1 G_{\nu\sigma}^2  
+ S_2(k_1,k_2) k_1^{\mu}G^1_{\mu\rho}G^2_{\nu\sigma}k_2^{\nu}
\right ],
\label{ff0++} \\
\epsilon_{\mu}T^{\mu \nu}\epsilon'_{\nu}
[2^{++}] & = & \epsilon_{\rho\sigma} 
 \left [ T_1(k_1,k_2)G^1_{\mu\rho}G^2_{\mu\sigma}  
 + T_2(k_1,k_2) k_1^{\rho}k_2^{\sigma}
 G_{\mu\nu}^1G_{\mu\nu}^2 \right. + 
\nonumber \\
& &  T_3(k_1,k_2) k_1^{\mu}G^{1}_{\mu\rho} G^{2}_{\nu\sigma}k_2^{\nu} 
\left. + T_4(k_1,k_2) k_1^{\rho}k_2^{\sigma}
k_1^{\mu}G^1_{\mu\rho}G^2_{\nu\rho}k_2^{\nu}\right ], 
\label{ff2++}
\end{eqnarray}
where 
\begin{equation}\label{pmunu}
P_{\rho\sigma} \equiv g_{\rho\sigma}-\frac {P_{\rho}P_{\sigma}}{m^2},
\end{equation}
for a resonance with mass $m$ and momentum $P_{\mu}$.  
Here $\epsilon_{\rho\sigma}$ is the polarization tensor
satisfying
\begin{eqnarray}\label{vector}
\sum_{\epsilon}
\epsilon_{\rho\sigma}\epsilon_{\rho^\prime\sigma^\prime} &
= & \frac
12(P_{\rho\rho^\prime}P_{\sigma\sigma^\prime}+P_{\rho\sigma^\prime}
P_{\sigma\rho^\prime})-\frac
13P_{\rho\sigma}P_{\rho^\prime\sigma^\prime}.
\end{eqnarray}

The {\it number} of form factors 
reflects the number of independent helicity amplitudes for the 
$\gamma \gamma$ where, for transverse (T) or longitudinally 
polarised (L) photons one forms

\begin{eqnarray}\label{helicity}
&& 0^{++}: \lambda = 0; TT or LL \nonumber \\
&& 0^{-+}: \lambda = 0; TT \nonumber \\
&& 1^{++}: \lambda = 0; TT:  \lambda= \pm 1; TL \nonumber \\
&& 2^{++}: \lambda = 0; TT , LL: \lambda =\pm 1; TL: \lambda =\pm 2; TT.
\end{eqnarray}
The {\it functional forms}
 of the $F_i(k_1,k_2)$ depend on the composition of $R$.
 In the particular case where the form factor is
modelled\cite{korner,koller,kuhn} as a QCD 
analogue\cite{barb} of the two photon coupling to 
positronium\cite{ale}, the various $F_1 \neq 0$ 
while $F_{2,3,4} = 0$: this has been discussed in ref\cite{cfl}. 
In this case there arise specific relations among
the helicity amplitudes which is the source of the polarisation
for the $1^{++}$ in eq.\ref{pol}. In the NRQM approximation\cite{kuhn}
\begin{eqnarray}\label{qqbar0++}
\epsilon_{\mu}T^{\mu \nu}\epsilon'_{\nu}
[0^{++}(q\bar q)]=c^\prime \sqrt{\frac 16}
 \left [ G_{\mu\nu}^a G_{\mu\nu}^a(m^2+k_1\cdot
k_2)-2 k_1^{\nu}G_{\mu\nu}^a G_{\mu\rho}^a k_2^{\rho}\right ]
/(k_1\cdot k_2)^2, \\
\equiv 
c^\prime m^2\sqrt{\frac 23}
  G_{\alpha\mu}^{1a} G_{\alpha\nu}^{2a}P_{\mu\nu}  
/(k_1\cdot k_2)^2,
\end{eqnarray}
and 
\begin{equation}\label{17}
\epsilon_{\mu}T^{\mu \nu}\epsilon'_{\nu}
[2^{++}(q\bar q)]=c^\prime \sqrt{2}m^2 G_{\mu\rho}^a G_{\nu\rho}^a
e^{\mu\nu}/(k_1\cdot k_2)^2,
\end{equation}
where the constants $c^\prime$ are proportional to the derivative of
the radial wavefunctions at the origin:
\begin{equation}\label{18}
c^\prime=g_s^2\sqrt{\frac 1{m^3\pi}} R^\prime(0).
\end{equation}
This structure implies that $2^{++}$ $^3P_2$ $q\bar{q}$ will be produced
polarised with the $\lambda=  0$ in the sense of
eqns.(\ref{pol} and \ref{helicity}) suppressed at$O(tt'/m^2)$.
This selection rule is expected to be realised even in the more 
physically relevant limit of light quarks\cite{bcl}.

The form factor for $0^{++}$ and $2^{++}$ glueballs in ref.\cite{qsum}
can be considered a natural relativistic generalization of TE mode
glueballs in a cavity approximation such as the MIT bag
model and the production amplitude takes the form\cite{cfl}
\begin{eqnarray}\label{extra3}
\epsilon_{\mu}T^{\mu \nu}\epsilon'_{\nu}
[J^{++}(G)]
= {\cal P^{(J)}}_{\mu\nu}
 \frac {G^{1a}_{\mu\rho} G^{2a}_{\nu\rho}}{k_1\cdot k_2} F(k_1;k_2)
\end{eqnarray}
where ${\cal P^{(0)}}_{\mu\nu} \equiv P_{\mu\nu}/\sqrt{3}$ and
${\cal P^{(2)}}_{\mu\nu} \equiv \epsilon_{\mu\nu}$. The form factor
$F(k_1;k_2)$ is determined by the glueball
radial wavefunction common
to the $0^{++}$
and $2^{++}$ states, so that the relative magnitudes of their form
factors are fixed and the ensuing $dk_T$ dependences will be similar.
The behaviour of $0^{++}$ and $2^{++}$ $q\bar{q}$ also will be similar to one
another but in general will differ from those of the glueballs. We shall 
not speculate here on particular models for such form factors but address
some general features.

For $(0,2)^{++}$ in general,any difference
in the glueball and  $q\bar{q}$ production
will be driven by the form factors
which  are functions of
two variables. The large momentum transfer behaviour of 
$P-wave$ $q\bar{q}$ and $S-wave$ glueballs with $J^{PC} = (0,2)^{++}$
may be constrained by power counting arguments\cite{cfl,bf}.
  When $R$ is an $L=0$ bound state of two constituents, the
leading large $k_1 \cdot k_2$ behaviour of $F_1(k_1;k_2)$ is 
$\frac{1}{(k_1 \cdot k_2)} f(z)$ 
(where $z  \equiv \frac{(k_1-k_2)\cdot (k_1+k_2)}{k_1
\cdot k_2}$).  The $F_i(k_1;k_2)$ entering the production amplitudes
with additional factors of $k_{1,2}^{\mu}$ have correspondingly more rapid 
falloff. For $L=1$ systems at large $k_1 \cdot k_2$
one expects an additional $\mu^2/k_1
\cdot k_2$ suppression, 
where $\mu$ is a scale reflecting the
variation of the wavefunction at the origin. 

The behaviour of the $L=0$ and $L=1$ wavefunctions will
also be expected to differ in general as their internal relative 
momenta $dp_T \to 0$.
A suggestive model is if $dk_T$ correlates with the internal 
momentum $dp_T$ such that in the $L$-th partial wave
$$
F(k_1; k_2) \sim |dk_T|^L \times \psi(t_1;t_2)
$$
In such a situation
as $|dk_T| \rightarrow 0$,
$^3P_{0,2}$ $q\bar{q}$ states will be killed while $0^+, 2^+$
states controlled by $S$-waves (such as glueballs or strong 
coupling
to pairs of $0^-$ mesons in S-wave) would survive. It is an open
question whether the $dk_T \neq 0$ of the production mechanism is
transferred into the relative momentum of the composite meson's
wavefunction. In the NRQM of eqns.(\ref{qqbar0++}-\ref{18}) this
does not occur;  in the $t$-channel of
$gg \to q\bar{q}$ where $(k_1-k_2)^2 \leq m_q^2$,
the massive quark propagator
 that is implicit in the derivation of eqn.\ref{18} dilutes 
any such correlation. However, in the light quark limit there is the
possibility for the singular behaviour of non-perturbative 
propagators\cite{pennington} to cause a strong correlation between $dk_T$ and
 the internal (angular) momentum. This may be tested by measuring
the $dk_T \to 0$ dependence of $0^{++},2^{++},4^{++}$ in $e^+e^- \to
e^+e^-R$
and test if the $|dk_T|$ transmits to the wavefunction
giving a $|dk_T|^L$ dependence. Our suggested strategy is as follows.

The similarity between the observed properties of $0^{-+}$ $2^{-+}$ and 
$1^{++}$ production in $pp \rightarrow ppR$ and those calculated
for $e^+e^- \rightarrow e^+e^-R$ suggest that either diffractive
scattering is driven by a colour singlet state transforming as a
conserved vector current or/and that $gg \rightarrow R$
is the elemental process in the pomeron-pomeron interaction.
This needs to be tested quantitatively. If verified, we may extend 
the concept of ``stickiness"\cite{chan}. The recommended strategy
is to measure $F^R_{\gamma \gamma} (k_1;k_2)$ in
$e^+e^- \rightarrow e^+e^-R$ and compare with the analogous
$F^R_{gg(?)} (k_1;k_2)$ in $pp \rightarrow ppR$. 
 Observation of
an identical $k_1, k_2$ dependence in $pp \to ppR$ for the production of 
established 
$q\bar{q}$ states, such as $f_2(1270)$, would establish the
conserved vector current dynamics
of the double-pomeron production process. Conversely,
the appearance of prominent states in $pp \rightarrow ppR$ that 
are suppressed in $e^+e^- \rightarrow e^+e^-R$ (``sticky" 
states\cite{chan}) and thereby are glueball candidates, would
enable extraction of their
$F_i(k_1;k_2)$. Such information would enable
comparison of the production of $q\bar{q}$ states and 
the enigmatic states, thereby untangling their
structure and dynamics.

Our general conclusions are as follows.

(i) The observed suppression of $0^{-+}$,$2^{-+}$ and $1^{++}$ 
as  $dk_T \rightarrow 0$, and also polarisation of the $1^{++}$
will arise if the production mechanism involves conserved vector currents.

(ii) The production of $0^{++},2^{++}$ is richer. 
In these cases it will be the dynamical behaviour
of the form factors $F_i(k_1;k_2)$, and hence the internal 
dynamics of the resonance $R$,
that will determine the outcome. Thus there exists the possibility
that $q\bar{q}$ and glueball states may be distinguishable
in the $0^{++},2^{++}$ sectors. It is already clear that not all states 
of a given $J^{PC}$ behave the same; the established $q\bar{q}$ $^3P_2$
$f_2(1270;1525)$ disappear as $dk_T \rightarrow 0$ whereas
$f_2(1930)$ survives\cite{wa102,kirkhad97}. 
To investigate the source of this it is 
necessary to measure the various $F_i(k_1;k_2)$ and to compare
$F_{(\gamma \gamma)}(k_1;k_2)$ with $F_{gg(?)}(k_1;k_2)$.

I am indebted to R.Cahn, M.Diehl, A.Donnachie, 
G.Farrar, J.Forshaw, J.-M.Fr\`ere, P.V.Landshoff  and R.G.Roberts
for discussions and comments
and to A.Kirk and G.Gutierrez for information about their data.

\begin{figure}
\epsfxsize = \hsize
\epsffile{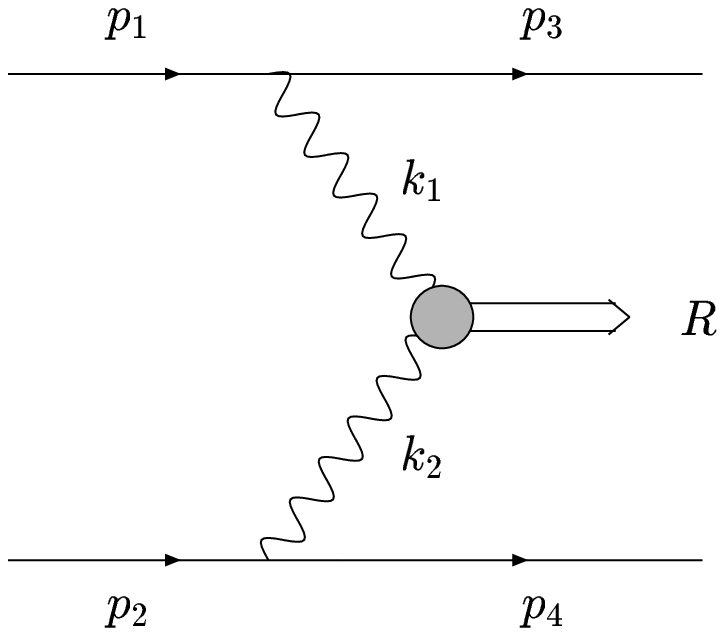}
\caption{Kinematics for central production of a resonance}
\label{a}
\end{figure}

\end{document}